\documentclass[prd,twocolumn,showpacs,preprintnumbers,amsmath,amssymb,superscriptaddress]{revtex4}
\usepackage{amsmath}
\usepackage{amssymb}
\usepackage{afterpage}
\usepackage{epsfig}
\usepackage{float}
\usepackage{rotating}
\usepackage{xspace}
\usepackage{dcolumn}
\usepackage{bm} 

\newcommand{\DFLive}{609.82}
\newcommand{\DRLive}{201.75}
\newcommand{\DFEvents}{28,994,380}
\newcommand{\DREvents}{8,898,551}

\begin{document}
\preprint{FERMILAB-PUB-07-134-E, BNL-78143-2007-JA, hep-ex 0705.3815}

\title{Measurement of the Atmospheric Muon Charge Ratio at TeV Energies with MINOS}         

\newcommand{\Cambridge}{Cavendish Laboratory, Univ. of Cambridge, Madingley Road, Cambridge CB3 0HE, UK}
\newcommand{\FNAL}{Fermi National Accelerator Laboratory, Batavia, IL 60510}
\newcommand{\RAL}{Rutherford Appleton Laboratory, Chilton, Didcot, Oxfordshire, OX11 0QX, UK}
\newcommand{\UCL}{Dept. of Physics and Astronomy, University College London, Gower Street, London WC1E 6BT, UK}
\newcommand{\Caltech}{Lauritsen Lab, California Institute of Technology, Pasadena, CA 91125}
\newcommand{\ANL}{Argonne National Laboratory, Argonne, IL 60439}
\newcommand{\Athens}{Department of Physics, University of Athens, GR-15771 Athens, Greece}
\newcommand{\NTUAthens}{Dept. of Physics, National Tech. Univ. of Athens, GR-15780 Athens, Greece}
\newcommand{\Benedictine}{Physics Dept., Benedictine University, Lisle, IL 60532}
\newcommand{\BNL}{Brookhaven National Laboratory, Upton, NY 11973}
\newcommand{\CdF}{APC -- Coll\`{e}ge de France, 11 Place Marcelin Berthelot, F-75231 Paris Cedex 05, France}
\newcommand{\Cleveland}{Cleveland Clinic, Cleveland, OH 44195}
\newcommand{\Delhi}{Dept. of Physics and Astrophysics, University of Delhi, Delhi 110007, India}
\newcommand{\GEHealth}{GE Healthcare, Florence SC 29501}
\newcommand{\DOE}{Div. of High Energy Physics, U.S. Dept. of Energy, Germantown, MD 20874}
\newcommand{\Harvard}{Department of Physics, Harvard University, Cambridge, MA 02138}
\newcommand{\HolyCross}{Holy Cross College, Notre Dame, IN 46556}
\newcommand{\IIT}{Physics Division, Illinois Institute of Technology, Chicago, IL 60616}
\newcommand{\Indiana}{Departments of Physics and Astronomy, Indiana University, Bloomington, IN 47405}
\newcommand{\ITEP}{High Energy Exp. Physics Dept., Inst. of Theor. and Exp. Physics, 
  B. Cheremushkinskaya, 25, 117218 Moscow, Russia}
\newcommand{\JMU}{Physics Dept., James Madison University, Harrisonburg, VA 22807}
\newcommand{\LASL}{Nucl. Nonprolif. Div., Threat Reduc. Dir., Los Alamos National Laboratory, Los Alamos, NM 87545}
\newcommand{\Lebedev}{Nuclear Physics Dept., Lebedev Physical Inst., Leninsky Prospect 53, 117924 Moscow, Russia}
\newcommand{\LLL}{Lawrence Livermore National Laboratory, Livermore, CA 94550}
\newcommand{\MIT}{Lincoln Laboratory, Massachusetts Institute of Technology, Lexington, MA 02420}
\newcommand{\Minnesota}{University of Minnesota, Minneapolis, MN 55455}
\newcommand{\Crookston}{Math, Science and Technology Dept., Univ. of Minnesota -- Crookston, Crookston, MN 56716}
\newcommand{\Duluth}{Dept. of Physics, Univ. of Minnesota -- Duluth, Duluth, MN 55812}
\newcommand{\Oxford}{Sub-dept. of Particle Physics, Univ. of Oxford,  Denys Wilkinson Bldg, Keble Road, Oxford OX1 3RH, UK}
\newcommand{\PSU}{Dept. of Physics, Pennsylvania State Univ., University Park, PA 16802}
\newcommand{\Pittsburgh}{Dept. of Physics and Astronomy, Univ. of Pittsburgh, Pittsburgh, PA 15260}
\newcommand{\IHEP}{Inst. for High Energy Physics, Protvino, Moscow Region RU-140284, Russia}
\newcommand{\RoyalH}{Physics Dept., Royal Holloway, Univ. of London, Egham, Surrey, TW20 0EX, UK}
\newcommand{\Carolina}{Dept. of Physics and Astronomy, Univ. of South Carolina, Columbia, SC 29208}
\newcommand{\SLAC}{Stanford Linear Accelerator Center, Stanford, CA 94309}
\newcommand{\Stanford}{Department of Physics, Stanford University, Stanford, CA 94305}
\newcommand{\Sussex}{Dept. of Physics and Astronomy, University of Sussex, Falmer, Brighton BN1 9QH, UK}
\newcommand{\TexasAM}{Physics Dept., Texas A\&M Univ., College Station, TX 77843}
\newcommand{\Texas}{Dept. of Physics, Univ. of Texas, 1 University Station, Austin, TX 78712}
\newcommand{\TechX}{Tech-X Corp, Boulder, CO 80303}
\newcommand{\Tufts}{Physics Dept., Tufts University, Medford, MA 02155}
\newcommand{\UNICAMP}{Univ. Estadual de Campinas, IF-UNICAMP, CP 6165, 13083-970, Campinas, SP, Brazil}
\newcommand{\USP}{Inst. de F\'{i}sica, Univ. de S\~{a}o Paulo,  CP 66318, 05315-970, S\~{a}o Paulo, SP, Brazil}
\newcommand{\Washington}{Physics Dept., Western Washington Univ., Bellingham, WA 98225}
\newcommand{\WandM}{Dept. of Physics, College of William \& Mary, Williamsburg, VA 23187}
\newcommand{\Wisconsin}{Physics Dept., Univ. of Wisconsin, Madison, WI 53706}
\newcommand{\deceased}{Deceased.}

\author{P.~Adamson}
\affiliation{\FNAL}
\affiliation{\UCL}

\author{C.~Andreopoulos}
\affiliation{\RAL}

\author{K.~E.~Arms}
\affiliation{\Minnesota}

\author{R.~Armstrong}
\affiliation{\Indiana}

\author{D.~J.~Auty}
\affiliation{\Sussex}

\author{S.~Avvakumov}
\affiliation{\Stanford}

\author{D.~S.~Ayres}
\affiliation{\ANL}

\author{B.~Baller}
\affiliation{\FNAL}

\author{B.~Barish}
\affiliation{\Caltech}

\author{P.~D.~Barnes~Jr.}
\affiliation{\LLL}

% ...

\collaboration{The MINOS Collaboration}
\noaffiliation

\date{\today}      

\begin{abstract}
The 5.4 kton MINOS far detector has been taking charge-separated cosmic ray muon data since the beginning of August, 2003 at a depth of 2070 meters-water-equivalent in the Soudan Underground Laboratory, Minnesota, USA. The data with both forward and reversed magnetic field running configurations were combined to minimize systematic errors in the determination of the underground muon charge ratio.   When averaged, two independent analyses find the charge ratio underground to be 
\begin{equation}
\label{chargeRatioUndergroundAbs}
N_{\mu^+}/N_{\mu^-} = 1.374 \pm 0.004\,({\rm stat.}) ^{+0.012}_{-0.010}\, ({\rm sys.}). \nonumber
\end{equation}
Using the map of the Soudan rock overburden, the muon momenta as measured underground were projected to the corresponding values at the surface in the energy range 1-7 TeV.  Within this range of energies at the surface, the MINOS data are consistent with the charge ratio being energy independent at the two standard deviation level.  
When the MINOS results are compared with measurements at lower energies, a clear rise in the charge ratio in the energy range 0.3 -- 1.0 TeV is apparent.  A qualitative model shows that the rise is consistent with an increasing contribution of kaon decays to the muon charge ratio.  
\end{abstract}

\maketitle

\section{Introduction}

The MINOS far detector is a 5.4 kton calorimeter, with magnetized steel planes located at the Soudan Underground Laboratory, Minnesota, USA.  It was designed to study neutrino oscillations with the NuMI beam which originates 735 km away at Fermilab.  At a depth of 710 meters below the Earth's surface, the MINOS far detector is the deepest experiment to measure cosmic ray muons with a magnetized detector,  thus providing a capability to distinguish $\mu^+$ from $\mu^-$ with large statistics.  The data correspond to muon energies at the surface in excess of approximately 1 TeV.  Above this energy, there are few  measurements of the charge ratio $N_{\mu^+}/N_{\mu^-}$.

Cosmic ray muons are produced when primary cosmic ray nuclei, mostly single protons, interact at the top of the atmosphere to produce hadronic showers.  The pions and kaons in these showers decay to muons, which are measured in detectors on the surface and underground, and with balloon-borne experiments in the atmosphere.  Since the cosmic ray primaries are positively charged, there are more positive than negative pions and kaons in the resulting hadronic showers.  In the energy 
range from 3 to 100 GeV, the CORT (Cosmic-Origin Radiation Transport) cosmic ray Monte Carlo~\cite{cort} predicts a relatively constant charge ratio of $N_{\mu^+}/N_{\mu^-} \approx 1.3$. In a compilation of measurements, the charge ratio was approximately constant at $N_{\mu^+}/N_{\mu^-} \approx 1.27$
with uncertainties increasing from $\sim1$\% at a few hundred MeV to $\sim6$\% at 300 GeV \cite{compile}.  A more recent measurement by the L3+C experiment at CERN  found similar results in the range 20 - 500 GeV \cite{l3}. 

At energies greater than a few hundred GeV, several competing processes can affect the charge ratio.  
The muons seen by MINOS result from pions and kaons that decay before they interact in the atmosphere.  As energy increases, the fraction of muons seen from kaon decays also increases because the longer-lived pions have become more likely to interact before decaying than the shorter-lived kaons.  Consequently, kaon decays begin to make an increasingly more important contribution to the muon charge ratio at these energies.  Since strong interaction production channels lead to a muon charge ratio from kaon decays that is greater than that from pion decays,
the measured charge ratio is expected to increase.  Several competing processes, however, could counter this increase at even higher energies.  Decays of charmed hadrons are one such process.  There is also the possibility that heavier elements become a more important component of cosmic ray primaries as the energy increases. This increasingly heavy composition would decrease the ratio of primary protons to neutrons, thereby decreasing the muon charge ratio. 
With careful measurements of the $N_{\mu^+}/N_{\mu^-}$ ratio in the cosmic rays, models of the interactions of cosmic rays in the atmosphere  can be improved.  

In addition, measurements of the cosmic ray muon charge ratio from a few GeV to a few TeV are important to constrain calculations of the atmospheric neutrino fluxes.  These are of interest both for detailed measurements of neutrino oscillations in atmospheric neutrino experiments and also for calculations of backgrounds for neutrino telescopes.  The muon charge ratio is a
particularly useful tool for testing the predicted atmospheric $\bar{\nu}/\nu$ ratio \cite{cort,honda,target}.

In MINOS, underground charge-separated cosmic muons were first studied in detail by Rebel \cite{rebel} and Beall \cite{beall}.

\section{The MINOS Far Detector}

\begin{figure}
\centerline{\epsfig{file=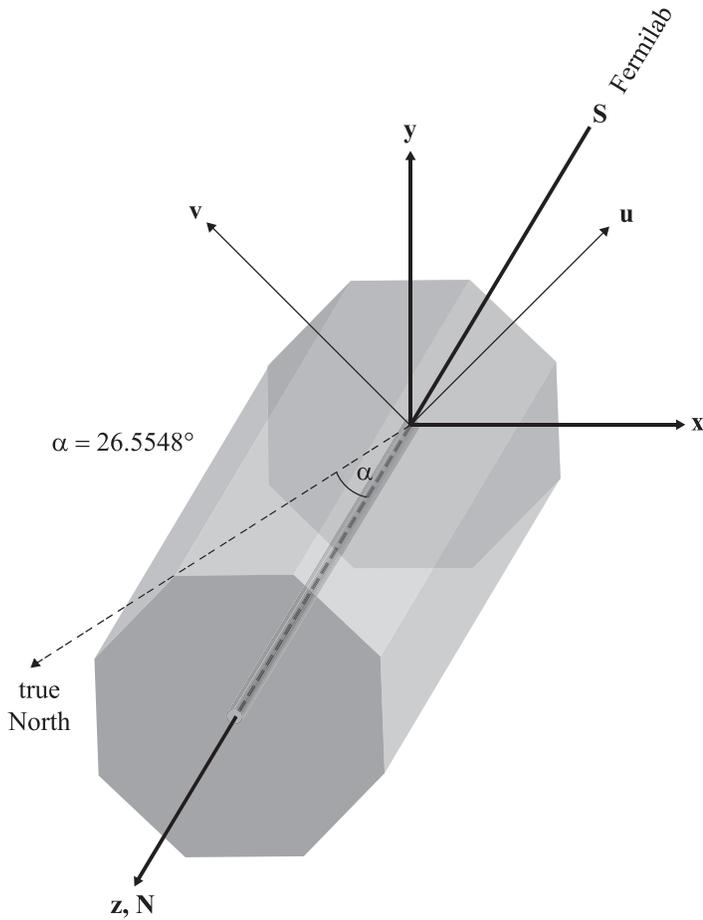,width=3.65in}}
\caption{\label{fig:minos}
The coordinate system for the MINOS cosmic ray analysis.  The octagonal steel and scintillator planes of the MINOS far detector are 8 m across.  The detector is 30 m long.  The central hole is for the magnet coil. The $+z$-axis is along the long axis of the detector and points toward detector North (N). Detector South (S) points back towards Fermilab.  The $y$-axis is directed toward the zenith.  The $x$-axis direction is chosen to make a right-handed coordinate system.  The origin of the coordinate system is the center of the South face of the detector.  Detector North (N) is rotated from true North by an angle $\alpha= 26.5548^\circ$ about the $y$-axis as measured by a gyro-theodolite; detector North therefore points along an azimuthal angle of $333.4452^\circ$.  Alternating planes of scintillator strips are oriented along either the  $u$ or $v$ axis directions, a coordinate system in which the $x-y$ plane is rotated by $+45^\circ$ about the $z$-axis.}
\end{figure}

The MINOS far detector is a steel-scintillator sampling and tracking calorimeter located at a depth of 2070 meters-water-equivalent (m.w.e.) in the Soudan Underground Laboratory in an iron mine in northern Minnesota (47.82027$^\circ$ N latitude, and 92.24141$^\circ$ W longitude)\cite{minosD}.
The detector is made of 486 octagonal planes of 2.54~cm thick steel laminates, interleaved with 484 planes of 1~cm thick extruded polystyrene scintillator strips at a 5.94~cm pitch.  
Each scintillator plane has 192 strips of width 4.1~cm.  The length of each strip depends on 
its position in the plane and varies between $3.4 - 8.0$~m.  The scintillator strips in alternating detector planes are oriented at $\pm 45^\circ$ to the vertical. The modular detector consists of two supermodules (SM) separated by a gap of 1.1~m. Fig.~\ref{fig:minos} shows the coordinate system used in the MINOS far detector cosmic ray analysis.  In terms of this coordinate system, events are described as coming from the zenith angle $\theta$ (the polar angle measured from the $y$-axis) from vertical ($\theta = 0^\circ$) toward the horizon ($\theta = 90^\circ$), and the azimuthal angle $\phi$ measured in the $x-z$ plane from true north ($\phi = 0^\circ$) to the east ($\phi = 90^\circ$).

Light from charged particles traversing the MINOS plastic scintillator is collected with wavelength shifting (WLS) plastic fibers embedded within the scintillator strips.  The WLS fibers are coupled to clear optical fibers at both ends of a strip and are read out using 16-pixel multi-anode phototmultiplier tubes (PMTs).  The signals from eight strips, each one of which is separated by approximately 1~m within the same plane, are optically summed and read out by a single PMT pixel.  The fibers summed on each pixel are different for the two sides of the detector, which enables the resulting eightfold ambiguity to be resolved for single particle events.  For all other types of events, ambiguities are effectively resolved using additional information from timing and event topology.  

The data acquisition and trigger have been described in \cite{atmo}.  The primary trigger requires 
activity to be observed on 4 planes out of 5 within 156~ns.  The detector calibration has been described in \cite{minosP}.  More detailed detector information can be found in \cite{minosD}. 

\begin{figure}
\centerline{\epsfig{file=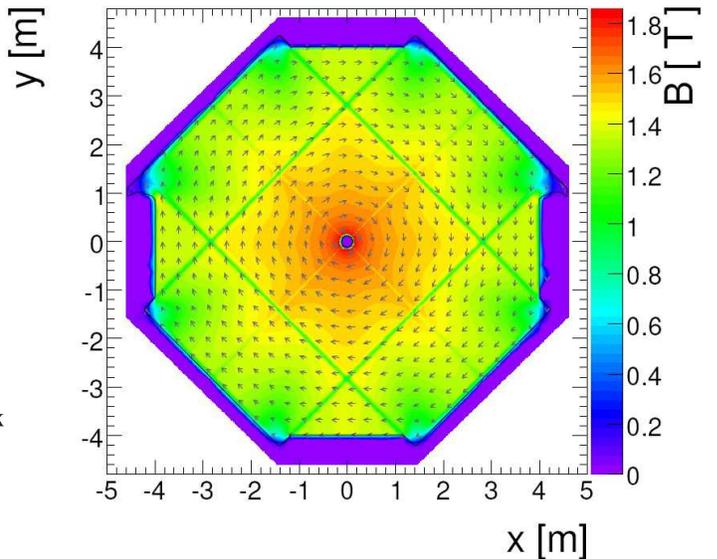,width=3.65in}}
\caption{\label{fig:map}
Finite element analysis model for the toroidal magnetic field in a plane of MINOS steel.  The coordinate system for the field map is shown in Fig.~\ref{fig:minos}.  In this map, the detector plane is being viewed from detector North.
}
\end{figure}

In order to measure the momentum of muons traversing the detector, the steel has been magnetized into a toroidal field configuration.  A finite element analysis calculation of the magnetic field strength for a typical MINOS detector plane is shown in Fig.~\ref{fig:map}.  These calculations show that each SM is magnetized to an average value of 1.3~T by the 15~kA current loop that runs through the coil hole (c.f., Fig.~\ref{fig:minos}) along the detector's $z$-axis.  The field is saturated near the coil hole at a strength of approximately 1.8~T, falling to almost 1~T near the edges.  There are small variations in field strength near the corners and along the gaps between the eight plates that make up each steel plane.  The detector field was designed to bend or `focus' negatively charged muons travelling from detector South (i.e., $\mu^-$ resulting from $\nu_{\mu}$ interactions in the detector from neutrinos 
originating in the Fermilab NuMI beam) toward the center of the detector. 

The toroidal magnetic field of the MINOS far detector impacts the acceptance of $\mu^{-}$ and $\mu^{+}$ entering the detector as a function of their incoming trajectories and the field direction. 
A muon is ``focused'' when the magnetic field steers it toward the center of the detector and ``defocused'' when directed away from the center.  These effects are most apparent for muons with trajectories that are parallel to the detector $z$ axis.  In one field orientation, Forward Field running, $\mu^{-}$ which enter the detector from the south and the $\mu^{+}$ which enter from the north will be focused into the center of the detector, while the muons in the opposite charge sign and trajectory combinations with be defocused.  Forward Field running  (``DF'' -- data forward) is the default configuration for MINOS data-taking with the NuMI beam from Fermilab.   MINOS has a second field orientation, Reverse Field running (``DR'' -- data reverse), in which  the coil current is reversed and $\mu^+$ from the south are focused into the detector.  These focusing/defocusing effects are most important at the edges of the detector acceptance and as a result, the charge ratio for muons with incoming trajectories on the edges of the detector acceptance will be either enhanced or suppressed depending on the charge and incoming direction of the muons.  

\section{Data Analysis}

\subsection{Cosmic Muon Data Sample}

In this paper we present results from data recorded between August 1, 2003 and February 28, 2006.  
During this period, the detector ran with both the DF and DR magnetic field configurations.   There were $\DFLive$ live days of Forward Field running and $\DRLive$ live days of Reverse Field running.   There were $\DFEvents$ events in the DF sample and $\DREvents$ events in the DR sample.  

\subsection{Event Selection}

The first stage of the event selection is to identify and remove periods of data associated with detector hardware problems.  The criteria defining bad runs are described in \cite{blake}.  The sample was selected using a series of cuts that are described in detail below and the numbers of events remaining at each stage in the selection are listed in Table~\ref{table:cutsummary}.  

\subsubsection{Pre-Analysis Cuts}
  
The first cut in the event selection requires at least one reconstructed track in the event (``1.  no reconstruction''). This  requirement  predominantly removes noise where the primary trigger was satisfied, but there was not enough activity to resolve the eightfold ambiguity from the optical summing of the scintillator strips.  The second requires that there is only a single track found by the track-fitting algorithm (``2.  multiples cut'').  The third requires that the coil be on and in a known state (``3.  coil status cut'').

\subsubsection{Analysis Cuts}

\begin{figure}
\centerline{\epsfig{file=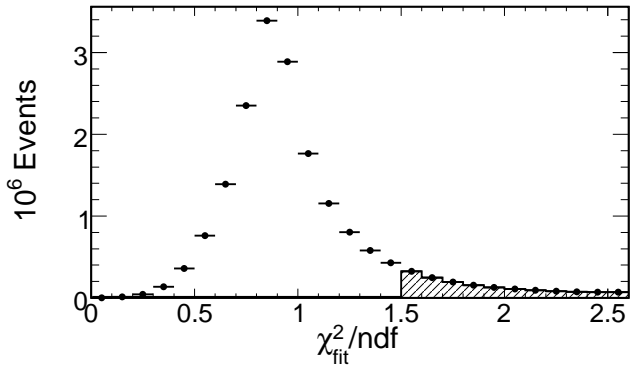,width=3.75in}}
\caption{\label{fig:chisq}
Distribution of $\chi^2_{fit}/ndf$ for the fits to cosmic muon tracks.  The cut was made at $\chi^2_{fit}/ndf > 1.5$.  The shaded region shows the excluded tracks. 
}
\end{figure}

The next set of cuts are meant to separate muon tracks from the background with high reliability.  These cuts require that: a track must cross at least 20 planes in the detector (``1.  20 plane cut''); a track must have a path length of at least 2~m (``2.  2.0~m track length cut''); the entrance  point of a track was required to be less  than 50~cm from an outside surface of the detector (``3.  Fiducial cut''); and a track  must pass a quality cut based on  $\chi^{2}_{fit}/ndf<1.5~$ (4.  ``fit quality cut'').  The $\chi^{2}_{fit}/ndf$ parameter is returned by the Kalman filter~\cite{fruhwirth}, which is the track fitting algorithm used in this analysis~\cite{minosD}.  The distribution used to select this cut value is shown in Fig.~\ref{fig:chisq} and this cut assures that the track found is a good fit to the track hit points. 

\begin{figure}
\centerline{\epsfig{file= 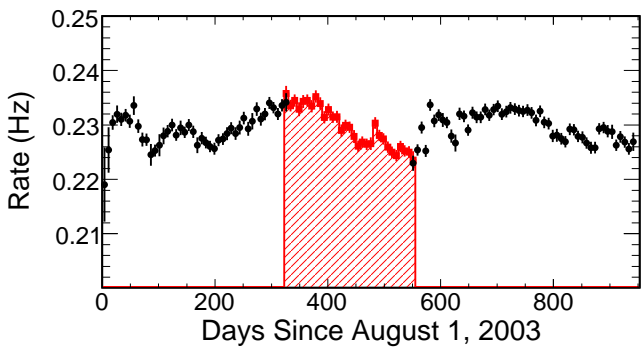,width=3.75in}}
\caption{\label{fig:rateVsDay}
Muon rates per day as a function of day number from the start of data taking with the complete and magnetized detector after the pre-analysis cuts  and analysis cuts 1-4.  The shaded area shows the period of reverse field running.
}
\end{figure}

In Fig.~\ref{fig:rateVsDay} we show the muon rate (Hz) as a function of day number from the beginning of data taking with the complete and magnetized detector after the pre-analysis cuts  and analysis cuts.  The fluctuations are consistent with seasonal variations in the cosmic ray muon flux \cite{seasonal}.

\subsubsection{Charge Sign Quality Cuts}

For the analysis presented in this paper it is necessary to cleanly identify the charge sign of selected muons and to ensure that systematic uncertainties in this identification are minimized.  This clean selection is achieved by placing charge-sign quality requirements on the reconstructed tracks. 
There are two components to the charge-sign quality cuts.  The first cut assures that the charge sign and momentum returned by the track-fitting algorithm has been well-determined.  
MINOS uses a Kalman filter technique for track-fitting~\cite{fruhwirth} that simultaneously determines which hits belong on a track and the momentum of the particle.  The technique involves a series of recursive matrix manipulations to specify the trajectory of the particle as well as the ratio of its charge to its momentum.  
The second cut minimizes residual systematic uncertainties that remain in the data set after the cuts already described have been applied.  To increase the robustness of the result two different approaches to identify a clean sample of events were adopted (2a,2b).  Consistent results for the cosmic ray muon charge-sign ratio were obtained with the two different approaches.  

\begin{enumerate}

 \item[(1)] ``track quality cut'' -- the significance of the measured muon charge sign and momentum for a track was required to be $(q/p)/\sigma_{q/p} \geq 2.2$, where $q$ is the charge sign and $p$ is the fit momentum.  Here $(q/p)$ is the fit parameter returned by the Kalman filter and the uncertainty on this fit parameter is $\sigma_{q/p}$.  For this cut we treat the quantity $(q/p)/\sigma_{q/p}$ as positive definite.  In Fig.~\ref{fig:rel_err} we show $N_{\mu^+}/N_{\mu^-}$, the ratio of the number of $\mu^+$ to $\mu^-$, as a function of $(q/p)/\sigma_{q/p}$ after all other cuts have been made since the final two cuts in the analysis are closely correlated.  For $(q/p)/\sigma_{q/p} \geq 2.2$, the charge ratio becomes asymptotically flat, suggesting that the charge sign and momentum are well fit.  As $(q/p)/\sigma_{q/p}$ tends to zero, the fitter is becoming less reliable at determining the charge sign.  For the lowest values of the significance, the fitter picks the two charge signs with equal probability and the measured charge sign ratio tends to unity.  Events with low values of the charge sign significance are typically high momentum tracks ($>~100$ GeV/c) that traverse the detector in such a way as to bend only slightly in the magnetic field.  These events have random charge sign identification.  
 
 \item[] Fig.~\ref{fig:rel_err} shows that the DF and DR data sets appear to behave very similarly with respect to the entire suite of charge sign quality cuts.  Not withstanding this similar behavior, there is a clear offset in the charge ratio on the plateau regions where the cuts define the tracks with well-determined charge sign.  For the DF data set, $N_{\mu^+}/N_{\mu^-} = 1.40$; for the DR data set, $N_{\mu^+}/N_{\mu^-} = 1.33$.  This overall systematic difference is discussed below.

\begin{figure}
\centerline{\epsfig{file=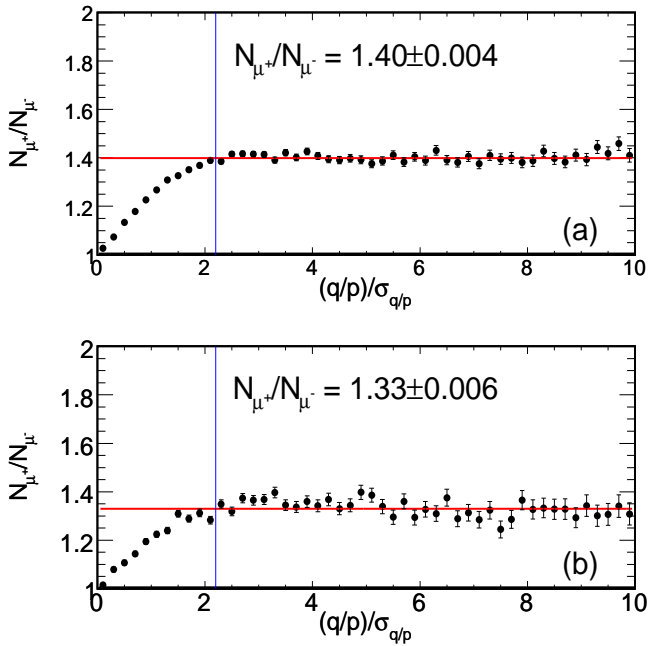,width=3.75in,height=3.5in}}
\caption{\label{fig:rel_err}
The $N_{\mu^+}/N_{\mu^-}$ ratio for reconstructed muon tracks as a function of $(q/p)/\sigma_{q/p}$ after all other cuts.  (a) Data Forward (DF) data distribution; (b) Data Reverse (DR) data distribution.  As indicated, the cut was chosen at $(q/p)/\sigma_{q/p} > 2.2$ for both data sets, where $(q/p)/\sigma_{q/p}$ becomes asymptotically flat.  The figures are labeled with the fits to a constant $N_{\mu^+}/N_{\mu^-}$ ratio for  the cut at $(q/p)/\sigma_{q/p} > 2.2$.   The fits are superposed onto the data as horizontal lines.
}
\end{figure}

\end{enumerate}

\noindent To check the consistency of the analysis, the final cut used was either (2a) or (2b).

\begin{enumerate}

\item[(2a)] {\it ``MIC''} (Minimum Information Cut) analysis -- a track was required to have at least 60 planes where the hit information was within 3.5 meters of the detector center; it was also required that the track fitter use the hits from these planes in its determination of $(q/p)$.  Here a `plane of hit information' is defined as a plane containing a strip from which signal is read out on both ends.  This cut was motivated by the need for high quality track information to resolve ambiguities in the charge sign determination for events with  $p \sim 50$ GeV/c.  

The effect of this cut is shown in Fig.~\ref{fig:mic}, where the muon charge ratio, $N_{\mu^+}/N_{\mu^-}$ is plotted for the (a) DF  and (b) DR data sets as a function of fit momentum. In this plot the number of planes of information required where the hits are within a radius of 3.5~m from the detector center has been varied from 0 (no cut) -100 planes.  Without the {\it MIC} cut, there is a large bump in the charge ratio distribution that peaks in the neighborhood of $40-50$ GeV/c and which decreases as the fraction of the track information in the inner part of the detector, where the field strength is  well-characterized, increases.  A second feature, which has much poorer statistics, appears near 100 GeV.  If these features were real physical effects, then we would not expect them to disappear as the quality of the fit information improved nor would we expect them to become reversed when the field was reversed.  But as is clear from Fig.~\ref{fig:mic}, the features do diminish as the number of planes of hit information increases and they do reverse when the field reverses.  Further, there is a large dip at low momenta in the DF data set that is not seen in the DR data set which also becomes negligible as the number of planes of hit information increases.  We therefore adopted the {\it MIC} cut at 60$-$planes to minimize these systematics.  The cut was placed at 60 planes because the addition of more planes of information did not appreciably reduce the systematic error while it does reduce the statistics for the number of events passing the cut.  
 
\begin{figure}
\centerline{\epsfig{file=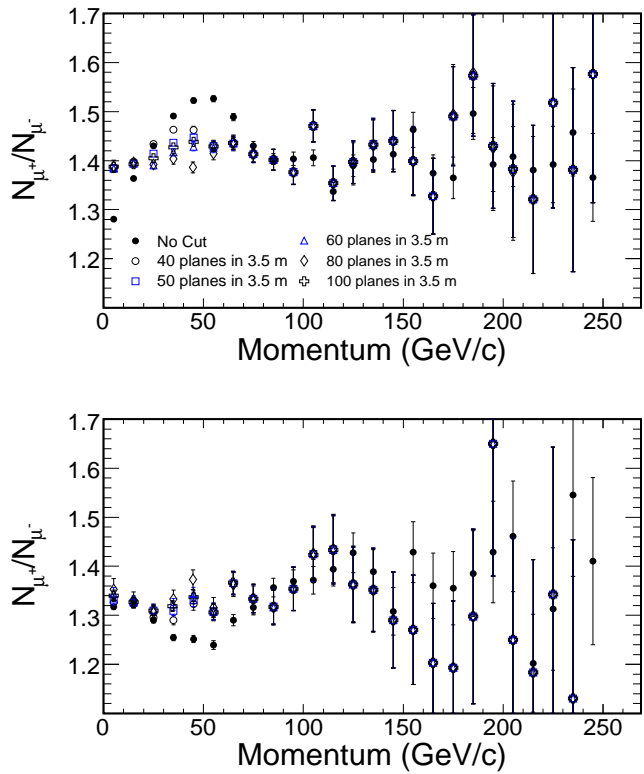,width=3.75in,height=4.25in}}
\caption{\label{fig:mic}
The muon charge ratio $N_{\mu^+}/N_{\mu^-}$ for (a) the DF and (b) the DR  data sets as a function of fit momentum as the number of planes of track information within 3.5~m of the detector center is varied from zero (no cut) - 100 planes.  
}
\end{figure}

\item[(2b)] {\it ``BdL''} analysis -- a track was required to have an integral field strength of  {\it BdL}  $> 12$ Tesla-meters (Tm), where {\it BdL} is a measure of  the perpendicular magnetic field $B_{perp}$ traversed by the track.  For this analysis  a variable {\it BdL} was defined as
\begin{equation}	
	BdL \equiv \int_{beg}^{end} B_{perp}(r)~{\rm d}L,
\label{eq:BdL}
\end{equation}   
where $B_{perp} = |\overrightarrow{B}(r) \times \overrightarrow{n}|$ is the component of the 
magnetic field perpendicular to the track direction $\overrightarrow{n}$ at a given point along the track path, $r$ is the distance from the detector center axis, ${\rm d}L$ is the differential pathlength element along the track, $beg$ is the point at which the muon enters the detector and $end$ is the point at which it either exits the detector or stops in the detector.  For our purposes, the track trajectory was approximated as a straight line running from the start to the end of the track.  Since the track length of a long track is 43\% in the steel, the cut at 12 Tm
corresponds to a $p_t$ kick from the magnetic field of 1.5 GeV.

\begin{figure}
\centerline{\epsfig{file=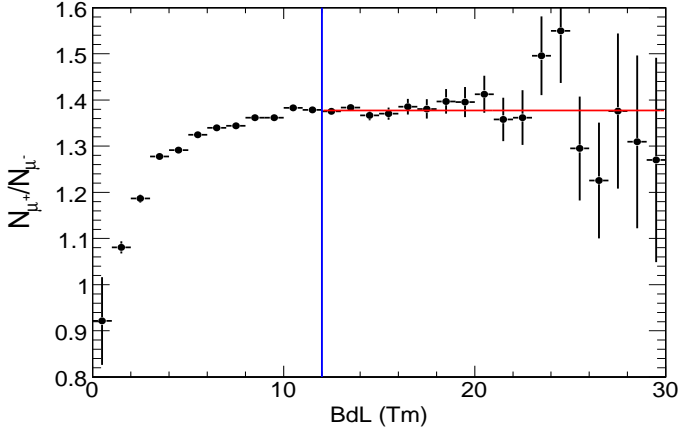,width=3.75in,height=2.45in}}
\caption{\label{fig:BdL}
The charge ratio $N_{\mu^+}/N_{\mu^-}$ as a function of {\it BdL}. The errors shown are statistical.  Superposed is the fit to a constant charge ratio $N_{\mu^+}/N_{\mu^-}$ for {\it BdL} $>$ 12.  
}
\end{figure}

The charge ratio as a function of the variable {\it BdL} is shown in Fig.~\ref{fig:BdL}.  For low values of {\it BdL}, the measured charge ratio approaches unity, as expected in the case of random charge determination.  As the integrated magnetic field increases, the charge ratio increases to a plateau value where the charge misidentification is highly suppressed. The plateau value is reached at BdL = 12~Tm, as shown in Fig.~\ref{fig:BdL}. Above this value the charge ratio as function of {\it BdL} is consistent with being constant. 
 
\end{enumerate}

\begin{table*}
\caption{\label{table:cutsummary} Summary of the Cuts Applied}

\begin{tabular}{|l|c|c|}

\hline \hline

 & ~{\bf DF$^a$} ~& ~{\bf DR$^b$}~ \\ \hline \hline
\# events before cuts
 &\multicolumn{1}{c|}{~N=$29.0\times10^{6}$~}&\multicolumn{1}{c|}{~N=$8.9\times10^{6}$~}\\ \hline \hline

\multicolumn{1}{|c|}{{\bf cut}}&\multicolumn{2}{|c|}{\bf Fraction Remaining}\\ \hline

  No Cuts & 1.0  & 1.0  \\ \hline

Pre-Analysis Cuts: &\multicolumn{2}{c|}{} \\ \hline

~~1. no reconstruction                       & 0.790 & 0.832 \\ \hline

~~2. multiples                                & 0.733  & 0.776  \\ \hline

~~3. coil status                              & 0.730  & 0.772  \\ \hline

Analysis Cuts: &\multicolumn{2}{c|}{} \\ \hline

~~1. 20 plane cut                                  & 0.554 & 0.585  \\ \hline

~~2. 2m track length cut                       & 0.551 & 0.582  \\ \hline

~~3. fiducial cut                                     & 0.534 & 0.565 \\ \hline

~~4. fit quality cut: $\chi_{fitter}^2$/{\it ndf} $<$ 1.5 & 0.427 & 0.452  \\ \hline

Charge-sign quality cuts &\multicolumn{2}{c|}{}   \\ \hline

~~1. $(q/p)/\sigma_{q/p} \geq 2.2$         & 0.141 & 0.147 \\ \hline

~~2a. {\it MIC} cut          & 0.048 & 0.050   \\ \hline

~~2b. {\it BdL} cut          & 0.033 & 0.031   \\ \hline

\hline \hline 

\multicolumn{2}{l}{$^a$ DF = cosmic data set, forward field}\\
\multicolumn{2}{l}{$^b$ DR = cosmic data set, reverse field}\\

\end{tabular}

\end{table*}

\noindent The effect of these cuts on the DF and DR data samples is given in Table~1.

As can be seen from Table~\ref{table:cutsummary}, tight cuts are required to minimize systematic errors in charge sign identification for the cosmic muon data sample.  Since the Forward (DF) and Reverse (DR) data samples behave similarly with respect to the total suite of cuts, we can use both data sets in subsequent analyses, even though there are overall systematic differences in their $N_{\mu^+}/N_{\mu^-}$ ratios. 

\section{Muon Charge Ratio at the MINOS Far Detector}

In the sections below we determine the muon charge ratio, $N_{\mu^+}/N_{\mu^-}$, as a function of  the reconstructed muon momentum, p$_{fit}$, by using the magnetic field of MINOS.   

\subsection{\label{sec:undergroundRatio}Measurement of the Muon Charge Ratio Underground}

\begin{figure}
\centerline{\epsfig{file= 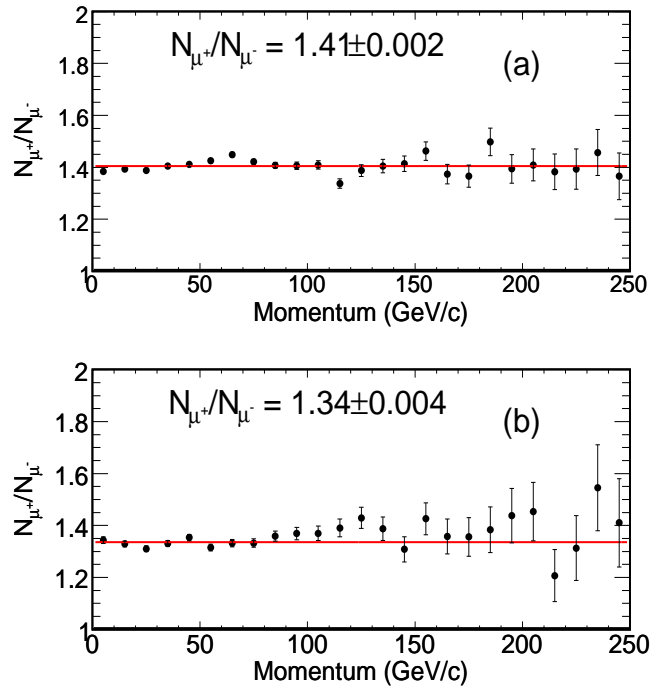,width=3.75in,height=3.75in}}
\caption{\label{fig:chargeRatioFR}
The charge ratio $N_{\mu^+}/N_{\mu^-}$ as a function of fit momentum 
for (a) the DF data set and (b) the DR data set.  For this figure, the data set was selected using the {\it MIC} cut.  Superposed on both are the fits to a constant charge ratio $N_{\mu^+}/N_{\mu^-}$.
}
\end{figure}

In Fig.~\ref{fig:chargeRatioFR} we show the charge ratio for events with reconstructed tracks that pass all cuts in Table~\ref{table:cutsummary} as a function of p$_{fit}$, the reconstructed momentum of the muon.  In this and the figures that follow, p$_{fit}$ is labeled ``Momentum''.  In Fig.~\ref{fig:chargeRatioFR} the charge ratio is shown separately for (a) the DF and (b) the DR data sets; for these distributions the {\it MIC} cut was used.  Superposed onto the data are fits to a constant charge ratio.  Fig.~\ref{fig:chargeRatioFR} clearly shows the systematic differences in the charge ratio measurements between Forward and Reverse field running.  

A method to cancel geometrical acceptance effects and alignment errors  
is discussed in \cite{matsuno,fields}.  If $A_1$ is the acceptance for $\mu^+$ and $A_2$ is the acceptance for $\mu^-$ in the Forward field direction, then the acceptances in the Reverse field direction are $A_1$ for $\mu^-$ and $A_2$ for $\mu^+$.
We can thus write two independent equations for the charge ratio in which the geometrical acceptances cancel: 
\begin{equation}
r_a =  (N_{\mu^+}  / t )_{DF}  /  (N_{\mu^-} / t )_{DR},
\end{equation}
\noindent and
\begin{equation}
r_b =  (N_{\mu^+} / t)_{DR}     /  (N_{\mu^-} /  t )_{DF},
\end{equation}
\noindent  By eliminating the Forward and Reverse live times, $t_{DF}$ and $t_{DR}$, between the two equations, we obtain a measurement of the mean charge ratio,~$r$, in which both geometrical acceptance and live time biases cancel,
\begin{equation}
\label{geomMean}
r = [r_a \times r_b]^{1/2}= [(N_{\mu^+}/N_{\mu^-})_{DF} \times (N_{\mu^+}/N_{\mu^-})_{DR}]^{1/2}.
\end{equation}
\noindent Eq.~(\ref{geomMean}) shows that it is the geometrical mean of the two independent charge ratio measurements that corrects for geometrical acceptance. 

Another class of systematic uncertainties, those that vary linearly with time, can also be cancelled by careful selection of the data analyzed.  If there are systematic effects that depend on live time in a linear way, these systematics can be cancelled by using Forward and Reverse data sets obtained during equal intervals of live time.  In this analysis we make two independent measurements of the charge ratio using data sets constructed in this way.  We first divide the Reverse data into two sets, $DR_1$ and $DR_2$, each of which has live time equal to $t_{DR}/2$.  We then pair $DR_1$ with $DF_1$, a Forward data set also with live time $t_{DR}/2$ which falls at the end of the first period of Forward running (c.f., Fig.~\ref{fig:rateVsDay}), in Eq.~(\ref{geomMean}).  These data result in one independent measurement of the charge ratio,~$r_1$, 
\begin{equation}
\label{eq:r1}
r_1 = [(N_{\mu^+}/N_{\mu^-})_{DF_1} \times (N_{\mu^+}/N_{\mu^-})_{DR_1}]^\frac{1}{2}
\end{equation}
\noindent The second measurement comes from pairing $DR_2$ with $DF_2$, a Forward data set with live time $t_{DR}/2$ which comes at the beginning of the second period of Forward running (c.f., Fig.~\ref{fig:rateVsDay}), in Eq.~(\ref{geomMean}).  These data result in second independent measurement of the charge ratio,~$r_2$,
\begin{equation}
\label{eq:r2}
r_2 = [(N_{\mu^+}/N_{\mu^-})_{DF_2} \times (N_{\mu^+}/N_{\mu^-})_{DR_2}]^\frac{1}{2}
\end{equation}
For these measurements, the data sets are all of length $t_{DR}/2$.   This analysis relies on systematic errors dominating over statistical errors since the method limits the sample sizes.  

In sections \ref{MIC} and \ref{BdL} the analysis just described is applied to the data sets generated with the {\it MIC} and {\it BdL} analyses.  

\subsection{\label{MIC} {\it MIC} Analysis}

\begin{figure}
\centerline{\epsfig{file= 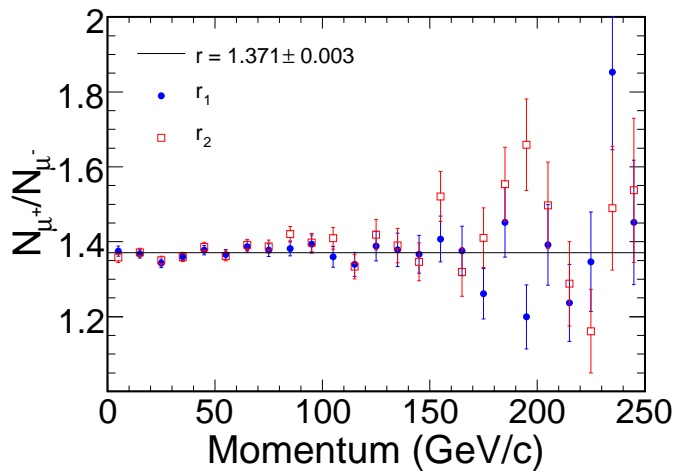,width=3.75in}}
\caption{\label{fig:money}The charge ratio $N_{\mu^+}/N_{\mu^-}$ as a function of fit momentum, p$_{fit}$. For this figure, the data set was generated using the {\it MIC} cut.  The errors shown are statistical.  Superposed is the fit to a constant charge ratio $N_{\mu^+}/N_{\mu^-}$.  }
\end{figure}

In Fig.~\ref{fig:money} we show the charge ratio $N_{\mu^+}/N_{\mu^-}$ for the two data sets $r_1$ and $r_2$ as a function of p$_{fit}$.  
The two independent measurements of the charge ratio are consistent with one another, which suggests we have significantly reduced the systematics seen in Fig.~\ref{fig:chargeRatioFR}.  We have used the data for $r_1$ and $r_2$ independently in a fit to a constant charge ratio over the range $0 \leq \text{p}_{fit} \leq 150$ (GeV/c).  In this fit, each fit momentum bin was assumed to have two independent measurements of the charge ratio.  The results of this fit give $N_{\mu^+}/N_{\mu^-} = 1.370 \pm 0.003$ with $\chi^2/ndf = 1.15$ for 29 degrees of freedom.  The best fit to a constant charge ratio has been superposed onto the $r_1$ and $r_2$ data in Fig.~\ref{fig:money} and is labeled $r$.

\begin{figure}
\centerline{\epsfig{file= 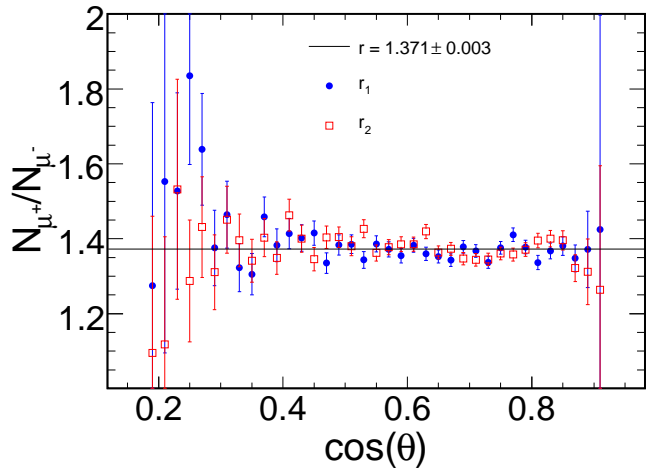,width=3.75in}} 
\caption{\label{fig:cosTheta}
The charge ratio $N_{\mu^+}/N_{\mu^-}$ as a function of $\cos{\theta}$, where $\theta$ is the zenith angle.  The errors shown are statistical.  Superposed is the fit to a constant charge ratio $N_{\mu^+}/N_{\mu^-}$.}
\end{figure}

\begin{figure}
\centerline{\epsfig{file= 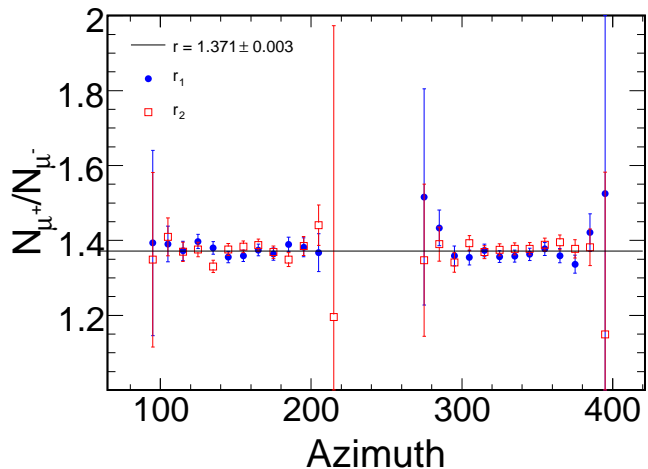,width=3.75in}}
\caption{\label{fig:az}
The charge ratio $N_{\mu^+}/N_{\mu^-}$ as a function of azimuthal angle $\phi$. The errors shown are statistical.   The gaps are due to acceptance effects resulting from the planar nature of the detector.  Muons with azimuths $<60^\circ$ have had $360^\circ$ added to their azimuth so that there is a continuous distribution of muons from the North ($270^\circ-420^\circ$).  Superposed is the fit to a constant charge ratio $N_{\mu^+}/N_{\mu^-}$. 
}
\end{figure}

In Figs.~\ref{fig:cosTheta}  and \ref{fig:az},  we show the charge ratio as a function of zenith angle and azimuth.  The analysis for both horizon coordinates follows the analysis for the charge ratio as a function of fit momentum.  There is no evidence for an angular dependence of the measured charge ratio.  

From Figs.~\ref{fig:money}, \ref{fig:cosTheta}, and \ref{fig:az},  we find small differences in the charge ratio depending on the parameter used to bin the data.  Some of these differences could result from the particular binning chosen for the different parameter plots.  To remove differences due 
to binning, we computed the charge ratios in Eq.~(\ref{eq:r1}) and Eq.~(\ref{eq:r2}) directly for all data with 
p$_{fit} < 250$ GeV/$c$.  Using the {\it MIC} selection in the charge ratio analysis, we find the bin free charge ratio is
\begin{equation}
\label{MICchargeRatio}
r = 1.372 \pm 0.003 ,
\end{equation}
where the error is statistical.  The number of events used in the computation of this ratio are given in Table~\ref{table: DFDR}.

\subsection{\label{BdL} {\it BdL} Analysis}

Using the {\it BdL} selection in the charge ratio analysis, we find the bin free charge ratio for all data with p$_{fit} < 250$ GeV/c is
\begin{equation}
\label{BdLchargeRatio}
r = 1.377 \pm 0.004 ,
\end{equation}
where the error is statistical.  The number of events used in the computation of this ratio are given in Table~\ref{table: DFDR}.

\begin{table}
\caption{\label{table: DFDR} Number of Events in {\it MIC} and {\it BdL} Samples}
\begin{tabular}{|l|c|c|c|c|}

\hline \hline

 \multicolumn{1}{|c}{ }&\multicolumn{2}{|c|}{{\it MIC}}&\multicolumn{2}{c|}{{\it BdL}}\\
\cline{2-5}
 \multicolumn{1}{|c}{ }& \multicolumn{1}{|c}{$N_{\mu^+}$ }& \multicolumn{1}{c|}{$N_{\mu^-}$} & \multicolumn{1}{c}{$N_{\mu^+}$} & \multicolumn{1}{c|}{$N_{\mu^-}$} \\
\cline{2-5}

~DF1~ & ~132,905~&~94,792~&~75,360~&~52,897~ \\
~DR1~ & 128,789&96,380&66,581&50,378 \\
~DF2~ & 133,382&94,434&73,889&51,126 \\
~DR2~ & 125,526&93,802&67,520&51,057 \\
\hline \hline

\end{tabular}

\end{table}

\subsection{\label{subsec:systUncert}Systematic Uncertainties}

Using the analysis technique described above many systematic uncertainties cancel.  However, there are residual systematic uncertainties which do not cancel.  These systematics can be  separated into two classes: bias errors and random charge identification errors.  Bias errors are those that lead to misidentifications of charge sign and cancel with a high degree of precision when combining forward and reverse field data.  The extent to which these errors do not cancel is a measure of the magnitude of these systematics.  Randomization errors lead to random misidentifications of charge that do not cancel in Eq.~(\ref{geomMean}).  The magnitude of these latter systematic errors have been determined by Monte Carlo simulation.  An example of a process that leads to randomization errors is multiple scattering, which can make a straight (high momentum) track appear curved to the track fitter.  The charge misidentification of these tracks is independent of charge sign.  A second source of randomization error comes from the inclusion of spurious hits in the track fit from bremsstrahlung, cross talk (signal appearing to come from a channel adjacent to the one hit), or the incorrect assignment of optically summed hits to a plane.  Since these hits are likely to fall on either side of the track with equal probability, the incorrect charge sign assignment again occurs with equal probability.

We examine three sources of systematic uncertainties.  The first two are bias errors and the third is due to randomization errors.  

\begin{enumerate}

\item {\it Combining Forward and Reverse Data.} 
Here we estimate the residual bias associated with focusing, detector acceptance, the magnetic field
map, and misalignments.   These cancel in principle when the forward and reverse data sets are combined with Eq.~(\ref{geomMean}).   As shown in Fig.~\ref{fig:money}, the method mostly suppresses the unphysical structures apparent in the charge ratio.  We take $(r_a - r_b)/2 = 0.009$, or the residual differences in the charge ratio due to incomplete cancellation of bias systematic errors remaining after the field has been reversed, as a measure of the systematic error associated with our procedure for combining forward and reverse data.  The data sets used to calculate $r_a$ and $r_b$ are the same as those used in the computation of $r_1$.

\item {\it Sliding Window.}
This bias error determines how stable our determination of the charge ratio remains with respect to time.  
As described in \ref{sec:undergroundRatio}, systematic errors which grow or decrease linearly in live time can be reduced by combining Forward and Reverse Field data sets in Eq.~(\ref{geomMean}) which have equal live times $t_{DR}/2$.  These determine the charge ratios $r_1$ and $r_2$.  We have recomputed $r_1$ and $r_2$ by successively sliding a window of width $t_{DR}/2$ to alternative times in the first and second periods of Forward Field running.  In this way we found a range of  $\pm 0.005$ for the charge ratio and this range was adopted as the systematic error.

\item {\it Random Charge Identification.} 
Random charge identification errors or `randomization' errors are those that lead to events getting random charge sign assignments.  Since the charge ratio $N_{\mu^+}/N_{\mu^-}$ is greater than one, more positive than negative muons will be misidentified and the measured charge ratio will decrease toward unity.  Or equivalently, randomization errors always result in a measurement of  the charge ratio
that is lower than its true value.  Consequently, randomization errors lead to a one-sided (positive) error on the measured charge ratio.

We estimated the magnitude of the randomization systematic error with Monte Carlo simulation and the {\it BdL} analysis.  We use the measurements of the charge ratio with successively tighter cuts on the {\it BdL} parameter (shown in Fig.~\ref{fig:BdL}), a process that systematically reduces the randomization error.  Using the trend from Monte Carlo in a $\chi^2$ fit, we estimated the charge ratio without randomization errors.  The difference between the fitted value of the charge ratio without randomization errors and the averaged (MIC/BdL) value of the charge ratio is 0.007.  We adopt this value for the systematic error due to random charge identification.  

\end{enumerate}

Table~\ref{tab:syst} summarizes these systematic uncertainties. The total systematic uncertainty was computed as the quadratic sum of individual uncertainties. 

\begin{table}[h]
\caption{\label{table: systematicErr}Systematic Uncertainties in the Charge Ratio}
\label{tab:syst}
\centerline{
\begin{tabular}{|l|c|}
\hline \hline
Source                             & $(\sigma_{syst})_i$   \\
\hline
1. Combining Forward/Reverse          & $\pm$ 0.009   \\
2. Sliding Window                     & $\pm$ 0.005   \\
3. Randomization              & +0.007     \\
\hline
$\sigma_{syst} = \sqrt{\sum_i(\sigma_{syst})_i^2}$                     & (+0.012, -0.010)   \\
\hline \hline
\end{tabular}
}
\end{table}

\subsection{\label{under} Muon Charge Ratio Underground}

For the charge ratio underground as measured by MINOS, we take the average of the charge ratio obtained from the {\it MIC} and {\it BdL} analyses, and use the systematic uncertainties listed in Table~\ref{tab:syst}, to give the muon charge ratio underground:

\begin{equation}
\label{chargeRatioUndergroundRes}
r = 1.374 \pm 0.004\,({\rm stat.}) ^{+0.012}_{-0.010}\, ({\rm sys.}).
\end{equation}

\section{\label{sec:CRProj}The Muon Charge Ratio at the Surface}

To infer the muon momentum at the surface from the momentum measured  underground requires knowledge of the rock overburden above the MINOS far  detector.  However, the rock overburden above the detector is of non-uniform composition with bands of iron formation embedded in Ely-Greenstone \cite{ruddick}. The topography of the surface above the MINOS far detector is also not level but rather has surface elevations that vary from 630 to 720 m over the angular region of interest \cite{kasa}.
Since the variations in the composition of the rock overburden are not known directly, the technique used here is to normalize the data to a ``world survey'' of vertical muon intensity data \cite{crouch}.   Thus, it is possible to derive a value of the slant depth for each solid angle bin.    

\subsection{\label{sec:Proj}Projection back to the Surface}

\subsubsection{\label{sec:measVertI}Measured Vertical Muon Intensity at the MINOS Far Detector}

In this analysis we use events with good charge sign and momentum reconstruction, that is, those events that pass the cuts in Table~\ref{table:cutsummary} up to and including the {\it MIC} cut.  As seen in Table~\ref{table: DFDR}, this data set has the largest statistics.  The selected events were first separated into bins of equal solid angle, $\Delta \Omega = \Delta \cos(\theta) \times \Delta \phi = 0.02 \times 0.10$ sr.  In each solid angle bin $j$, the vertical muon intensity was computed according to

\begin{equation}
\label{intensityI}
(I_\mu)_j  = \frac{1}{T} \frac{m N_j}{ \left(\epsilon_jA_j \Delta \Omega 
/\cos{\theta_j}\right)},
\end{equation}

\noindent where $T$ is the live-time; $N_j$ is the number of single muons in bin $j$; $m$ is a multiplicative factor, assumed independent of slant depth and direction, that accounts for muon multiplicity;  $\epsilon_j = \epsilon(\cos{\theta_j},\phi_j)$ and $A_j = A(\cos{\theta_j},\phi_j)$ are the efficiency and the projected area of the detector, respectively, as a function of zenith and azimuthal angles; and the $\cos(\theta_j)$ factor corrects for the muon intensity zenith angle
dependence at the surface to a good approximation in the momentum range
relevant here.
The intensity is converted to vertical intensity to facilitate comparison with the world survey data.
Each measurement of $(I_\mu)_j$ is an independent measurement of the vertical muon intensity in direction $(\cos{\theta_j},\phi_j)$.  

The efficiency of the far detector, $\epsilon_j = \epsilon(\cos{\theta_j},\phi_j)$, for the reconstruction of single muon tracks was computed with Monte Carlo generated muons.  
For this calculation, we generated a sample of $1.2 \times 10^6$ cosmic muons by Monte Carlo simulation.  Each event was generated by first choosing an arrival direction from the zenith angle dependence of the muon flux parameterization on the surface \cite{part} and then associating this direction with the overburden  [ g/cm$^2$] in the  Soudan~2 slant depth map  \cite{kasa}.  The energy of the event was selected from the surface cosmic ray muon distribution~\cite{part}.  With the energy and overburden, the event was tested to see whether it penetrated to the detector \cite{gaisser}.  The events that survived were placed on an imaginary box positioned around the detector \cite{Hawthorne:1998mz} and then propagated through the detector with the GEANT3 simulation of the MINOS detector.  

The standard method for computing the statistical error in the efficiency is to consider the application of the cuts to be a binomial process.  With the large number of events in our Monte Carlo sample, the statistical error is much less than 1\%.  However, our Monte Carlo does not include multiples, a fair sample of demultiplexing failures, or events arriving when the magnetic field is off and these effects outweigh the statistical errors.   Long experience with our MINOS Monte Carlo suggests that the uncertainty on our computation of $\epsilon$ is of the order of a few percent. 

The projected area of the MINOS far detector in direction ($\cos{\theta_j}, \, \phi_j$), $A_j = A(\cos{\theta}_j,\phi_j)$, was computed by first finding the unit vector along this direction, ${\bf \hat{n}_\mu}={\bf \hat{n}_\mu}(\cos{\theta_j}, \phi_j)$, and then defining the normal for each of the ten surfaces of the MINOS far detector, ${\bf\hat{n}_k}$, where $k=1-10$.  The projected area is then given by 

\begin{equation}
A_j = A(\cos{\theta}_j,\phi_j) = \sum_{k=1}^{10}({\bf \hat{n}_\mu}\cdot {\bf\hat{n}_k}) S_k,
\end{equation}

\noindent for all $({\bf \hat{n}_\mu}\cdot {\bf\hat{n}_k}) \geq 0$ and where $S_k$ is the area of the $k$th surface of the MINOS far detector.  Our computations show that the total acceptance for the MINOS far detector to single atmospheric muons is $(\epsilon A \Omega) = \sum_j (\epsilon_j A_j \Delta \Omega) = 1.3 \times 10^6$ cm$^2$ sr.    

To compare our results to those of other underground experiments \cite{part}, we make a correction to our measured vertical intensity to account for muon multiplicity, the correction factor $m$ in Eq.~(\ref{intensityI}).  Using a day of MINOS far detector data, we find that $m = 1.04$.  
Corrections were not made for the lateral distribution of multiple muons over the finite size of the MINOS detector.  These corrections are quite labor-intensive to compute and we conservatively estimate them  to be $\sim$10\%.   The effect of these uncertainties on the analysis are considered in \S\ref{sec:mcrs}.

\subsubsection{\label{sec:standardRock}Determination of the MINOS Slant Depth Map for Standard Rock}

For each solid angle bin $j$ in direction $(\cos{\theta_j},\phi_j)$ we computed a value for the slant depth, $X_j$, or the column of rock from the MINOS cavern to the surface in units of m.w.e.~(where 1 m.w.e.~= $10^2$ g/cm$^2$) by equating   our measured vertical muon intensity to Crouch's all-world average vertical muon intensity \cite{crouch}.   This parameterization represents the integral of muons of all energies at the surface that can reach the detector through a rock depth $X_j$.  In particular, we varied $X_j$ until the Crouch parameterization and our measured vertical intensities agreed.  Some solid angle bins at the edge of the acceptance had too few events to reliably determine the slant depth; these solid angle bins and their events were removed from further analysis.  One solid angle bin located near the zenith and on the edge of the detector acceptance had just enough events for the slant depth of that bin to be calculated.  However, the calculated slant depth was several hundred m.w.e. less than the slant depth of  neighboring solid angle bins.  Given the small acceptance of the bin and the peculiar slant depth derived for it, this solid angle bin and its events were also removed from the analysis.

\begin{figure}
\centerline{\epsfig{file= 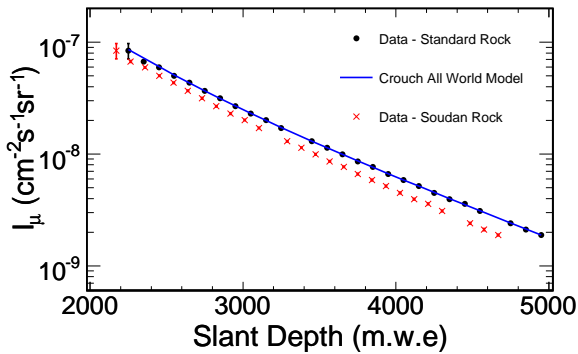,width=3.25in}}
\caption{\label{fig:vertIntensity}
The vertical muon intensity in the MINOS far detector hall.  Shown as solid circles are the vertical intensity data which have been normalized to the Crouch all-world average vertical intensity for standard rock  \cite{crouch}.  Overlaid onto the solid circles is the Crouch average.  Shown as x's is the vertical intensity for the MINOS far detector hall corrected for Soudan rock.
}
\end{figure}

Since the Crouch parameterization is given in terms of  ``standard rock'', the MINOS slant depth map computed here is in terms of standard rock.  In Fig.~\ref{fig:vertIntensity} we show the vertical muon intensity in the range $2000\,{\rm m.w.e.} \leq X \leq 5000 \,{\rm m.w.e.}$ made using the MINOS slant depth map with the Crouch parameterization superposed.  The two distributions coincide as they must. 

\subsubsection{\label{sec:soudanRock}Vertical Muon Intensity for Soudan Rock}

Muons of energy  $E_{\mu,0}$ at the Earth's surface lose energy \cite{part} as they traverse a slant depth $X$ through the Soudan rock to the MINOS detector according to 
\begin{equation}
\label{eloss}
-\frac{{\rm d}E_\mu}{{\rm d}X} = a(E_\mu)  + b(E_\mu) E_\mu,
\end{equation}
where the parameters $a$ and $b$ describe the energy lost by collisional and radiative processes, respectively.  Eq.~(\ref{eloss}) assumes continuous energy loss and does not account for fluctuations \cite{lipari}.  The energy loss parameters for standard rock, $(a_s, b_s)$,  as a function of energy are given in \cite{part}.  At the detector, the energy of the muons, $E_\mu$, can approximately be related to $E_{\mu,0}$ by \cite{part}
\begin{equation}
\label{surf} 
E_{\mu,0} = (E_\mu + a/b)e^{bX} - a/b.
\end{equation}
To convert the MINOS slant depths for standard rock to Soudan rock, we equate the minimum energy required to reach slant depth $X_s$ of standard rock to the minimum energy required to reach the equivalent slant depth $X_M$ for Soudan rock  \cite{meyer, menon, demuth}.  If $(a_M, b_M)$ describe the energy loss parameters for the Soudan rock at MINOS, then   
\begin{equation}
X_M = \frac{1}{b_M}\ln{[1+(\frac{a_s}{b_s})(\frac{b_M}{a_M}) (e^{b_s X_s} -1)]}.
\end{equation}
The values of the energy loss parameters $a$ and $b$ depend on the average composition of the rock \cite{demuth}.  For the collisional term,
\begin{equation}
a_M = a_s \langle\frac{Z_M}{A_M}\rangle \langle\frac{Z_s}{A_s}\rangle^{-1}, \nonumber
 \end{equation}
and for the radiative term
\begin{equation}
b_M = b_s\langle\frac{Z^2_M}{A_M}\rangle \langle\frac{Z^2_s}{A_s}\rangle^{-1}, \nonumber 
\end{equation}
where $Z$ is the atomic number and $A$ is the atomic mass.  The parameters for standard rock and the Ely-Greenstone rock at the Soudan Underground Laboratory  \cite{ruddick} are shown in Table~\ref{table: rockParam}.

\begin{table}
\caption{\label{table: rockParam} Rock Parameters}
\begin{tabular}{|l|c|c|}

\hline \hline

 &\multicolumn{1}{|c|}{~$\langle Z/A\rangle$~}&\multicolumn{1}{|c|}{~$\langle Z^2/A\rangle$~}\\
\hline 

Standard Rock & 0.5&5.5 \\
Soudan Rock & 0.5 & 6.1\\
\hline \hline

\end{tabular}

\end{table}

\begin{figure}
\centerline{\epsfig{file= 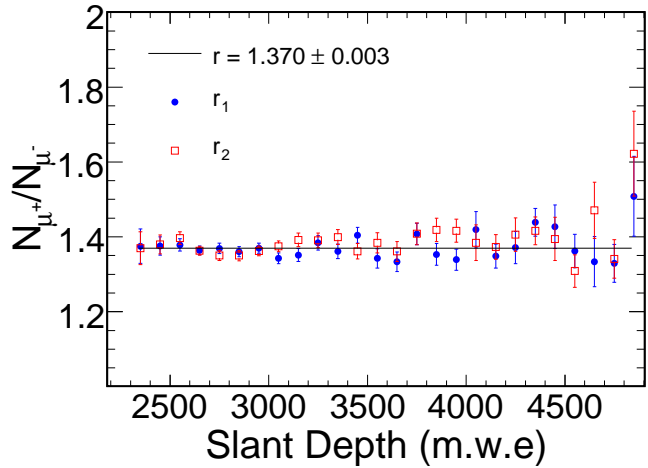,width=3.75in}}
\caption{\label{fig:slantDepthRatioOnly}  The charge ratio $N_{\mu^+}/N_{\mu^-}$ as a function of slant depth in Soudan rock.  The errors shown are statistical.  Superposed is the fit to a constant charge ratio $N_{\mu^+}/N_{\mu^-}$. }
\end{figure}

The vertical intensity using the Soudan rock map is shown in  Fig.~\ref{fig:vertIntensity}.  Since $\langle Z^2_M/A_M\rangle$ is larger for Soudan rock than for standard rock, a muon will lose more energy traversing an equivalent column of Soudan rock and therefore the vertical intensity at MINOS will have a steeper slope, as seen.  In Fig.~\ref{fig:slantDepthRatioOnly} we use the MINOS slant depth map to plot the charge ratio as a function of the slant depth.  Fig.~\ref{fig:slantDepthRatioOnly} shows that the charge ratio has little dependence on slant depth.

\subsubsection{\label{sec:proj}Projection back to the Surface}

\begin{figure}
\centerline{\epsfig{file= 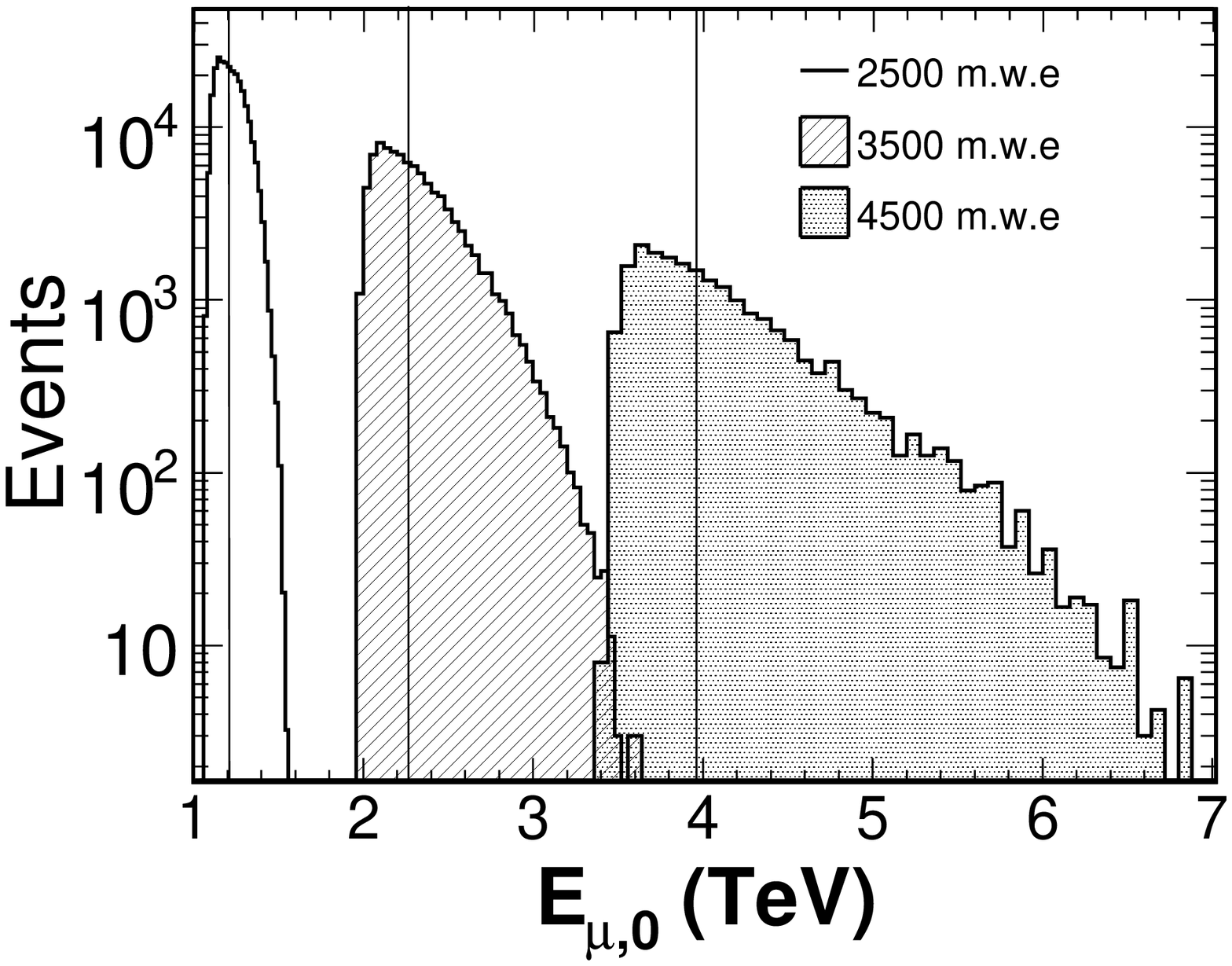,width=3.75in,height=2.68in}}
\caption{\label{fig:surfE} The distribution of muon energies projected back to the surface for three slant depth bins of width 100 m.w.e~used in the computation of the vertical intensity in Fig.~\ref{fig:vertIntensity}.  These projections were made with Eq.~(\ref{surf}) and Soudan rock parameters.  The median value of the muon energy $\langle E_{\mu,0}\rangle_{med}$ on the surface for these three bins is shown.
}
\end{figure}

The muon energy underground is obtained from the reconstructed momentum. To project the reconstructed muon energies underground, $E_\mu = c$p$_{fit}$, back to the surface, we used Eq.~(\ref{surf}) with Soudan rock parameters and the MINOS rock map.  In these projections, we assumed that the Soudan energy loss parameters varied with energy in a manner similar to the energy loss parameters for standard rock in \cite{part}.  In Fig.~\ref{fig:surfE} we show the result of this projection for three slant depth bins used to compute the vertical intensity in Fig.~\ref{fig:vertIntensity}.  This figure shows the distributions of surface muon energies, $E_{\mu,0}$, for these three bins, as well as the median energy, $\langle E_{\mu,0}\rangle_{med}$, for each distribution.  The width of the slant depth bins is 100 m.w.e.~and this width is the dominant contributor to the width of the surface energy distributions shown.  The expected increase of surface energy with increasing slant depth is clearly evident.

\subsection{\label{sec:mcrs}The Energy Dependence of the Muon Charge Ratio at the Surface}

Using Eq.~(\ref{surf}) we projected the muons from the {\it MIC} analysis back to the surface.  For each successfully reconstructed muon underground, we use its p$_{fit}$  value, its slant depth $X$, and the Soudan values of $(a_M, b_M)$ to obtain the surface energy $E_{\mu,0}$.   In these projections, we assumed that the Soudan energy loss parameters were independent of charge sign and they varied with energy in a manner similar to the energy loss parameters for standard rock in \cite{part}.

\begin{figure}
\centerline{\epsfig{file= 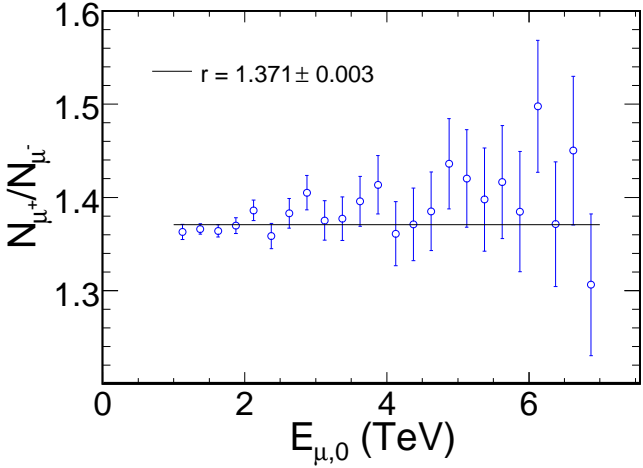,width=3.75in}}
\caption{\label{fig:ratioSurfE}   The muon charge ratio $N_{\mu^+}/N_{\mu^-}$ at the Earth's surface.  The errors shown are statistical.
}
\end{figure}

Once projected back to the surface, we sorted the muons into bins of width 0.25 TeV and then computed the charge ratio.  The results are shown in Fig.~\ref{fig:ratioSurfE}, where the charge ratio $N_{\mu^+}/N_{\mu^-}$ is plotted as a function of surface muon energy $E_{\mu,0}$.  In Table~\ref{table:surfCR} we give the charge ratio data in each energy bin, as well as the weighted mean.

\begin{table}
\caption{\label{table:surfCR} Charge Ratio at the Surface}

\begin{tabular}{|c|c|}
\hline \hline

{\bf $E_{\mu,0}$} (TeV)& {\bf $r$}  \\ \hline 

 1.00 -- 1.25& 1.363 $\pm$ 0.008 \\
 1.25 -- 1.50& 1.366 $\pm$ 0.006 \\
 1.50 -- 1.75& 1.364 $\pm$ 0.007 \\
 1.75 -- 2.00& 1.370 $\pm$ 0.009 \\
 2.00 -- 2.25& 1.386 $\pm$ 0.011 \\
 2.25 -- 2.50& 1.359 $\pm$ 0.014 \\
 2.50 -- 2.75& 1.383 $\pm$ 0.016 \\
 2.75 -- 3.00& 1.405 $\pm$ 0.019 \\
 3.00 -- 3.25& 1.375 $\pm$ 0.021 \\
 3.25 -- 3.50& 1.377 $\pm$ 0.023 \\
 3.50 -- 3.75& 1.396 $\pm$ 0.027 \\
 3.75 -- 4.00& 1.413 $\pm$ 0.031 \\
 4.00 -- 4.25& 1.361 $\pm$ 0.034 \\
 4.25 -- 4.50& 1.371 $\pm$ 0.039 \\
 4.50 -- 4.75& 1.385 $\pm$ 0.042 \\
 4.75 -- 5.00& 1.436 $\pm$ 0.048 \\
 5.00 -- 5.25& 1.420 $\pm$ 0.052 \\
 5.25 -- 5.50& 1.398 $\pm$ 0.055 \\
 5.50 -- 5.75& 1.417 $\pm$ 0.061 \\
 5.75 -- 6.00& 1.385 $\pm$ 0.065 \\
 6.00 -- 6.25& 1.498 $\pm$ 0.071 \\
 6.25 -- 6.50& 1.371 $\pm$ 0.067 \\
 6.50 -- 6.75& 1.450 $\pm$ 0.080 \\
 6.75 -- 7.00& 1.306 $\pm$ 0.076  \\

\hline \hline

$<r>$ & 1.371 $\pm$ 0.003 \\

\hline \hline

\end{tabular}

\end{table}

We have performed 1-parameter and 2-parameter fits to the data in Table~\ref{table:surfCR} over the surface muon energy range $1.0 \,\,{\rm TeV} < E_{\mu,0} < 7.0 \,\, {\rm TeV}$.  A fit to a constant charge ratio gives  
\begin{equation}
\label{chargeRatio}
N_{\mu^+}/N_{\mu^-} = 1.371 \pm 0.003\,({\rm stat.})  ^{+0.012}_{-0.010}\, ({\rm sys.}),
\end{equation}
with $\chi^2/ndf = 63.2/67$.  The data are consistent with a charge ratio that is independent of energy.

The 2-parameter linear fit gives 
\begin{equation}
N_{\mu^+}/N_{\mu^-} = (1.354 \pm 0.007)+ (0.85 \pm 0.33) \times 10^{-2}  E_{\mu,0}  ,
\label{eq:twoParam}
\end{equation}
where $E_{\mu,0}$ is in TeV and $\chi^2/ndf = 56.6/66$. 

There are two contributions to the systematic error on the slope:  uncertainties in the calculation of the energy scale $E_{\mu,0}$ and errors on the determination of the slope due to randomization.  Errors in the energy scale are mostly due to uncertainties in the slant depth map.   Using Monte Carlo methods to study these slant depth uncertainties, the errors are found to be $\sim$10\%.  Calculations show that errors in the surface muon energies resulting from systematic uncertainties in the slant depth map of this order are 15-20\% at 2100 m.w.e. and 25\% at 4000 m.w.e.
In a second test, the Soudan 2 slant depth map \cite{kasa} was substituted for the MINOS slant depth map and the surface energies recomputed.  The differences in the surface energies determined with these two maps are again approximately 20\%.  
We therefore estimate the systematic error on the energy scale to be $\pm$~20\%.  The uncertainty in the energy scale does not affect the significance of the slope.

Systematics due to randomization do affect the significance of the slope determination. For long track lengths, the tracks are better reconstructed, the charge misidentification is smaller, and the measured charge ratio systematically rises.  Thus randomization can mimic and cover up a rising dependency of the muon charge ratio on muon surface energy.   To estimate the size of this effect, we have separated the data into six subsets with {\it BdL} values from 10 Tm to 15 Tm.  As expected, we found these data sets to have different values of the charge ratio due to different but fixed amounts of randomization at the level of approximately 2\%.  In contrast, the values for the slope of the charge ratio versus surface energy were similar.  This randomization-free value for the slope parameter was 30\% smaller than the value from the two parameter fit.
We take this difference to be the systematic error on the slope,
\begin{equation}
slope = 0.85 \pm 0.33(stat) \pm 
0.26(syst) \times 10^{-2} \, {\rm TeV}^{-1}
\label{eq:slopesigma}
\end{equation}
Adding the errors in quadrature shows that the slope differs from zero by two sigma.  An alternative fit using a linear dependence on log $E_\mu$ yields a similar result.

\section{\label{sec:discussion}Discussion}

In order to obtain the charge ratio results shown in Fig.~\ref{fig:ratioSurfE}, each muon in our sample was projected back to the surface with Eq.~(\ref{surf}) using the Soudan energy loss parameters $(a_M,b_M)$ for both $\mu^+$ and $\mu^-$.  A theoretical complication to this procedure is the possibility that the energy loss parameters $a$ and $b$ are different for the two charges.  For ionization losses, it was pointed out by Fermi \cite{fermi} that the electrons in matter would introduce small differences in energy loss.  Calculations by Jackson and McCarthy \cite{jackson} confirmed that negative particles lose 
energy at a slower rate, with the difference dropping from tens of percent at MeV energies to about 0.3\% in the GeV range.  These calculations were subsequently verified experimentally both at MeV energies \cite{barkas1, barkas2, barkas3} and in the GeV range \cite{gev}. The approximations used in \cite{jackson}, however, break down when going to even higher energies and so more exact numerical methods are needed.  A calculation at TeV energies can be found in Jackson~\cite{jackson98} that shows a small 0.15\% increase in the ionization loss for $\mu^+$ over $\mu^-$.
For radiative energy loss, which involves the parameter $b$, the difference between $\mu^+$ and $\mu^-$ is much smaller and falls with energy \cite{ims}.  In the extrapolation used to obtain the results below, we assumed the same energy loss function for both charges.

The projections of our data back to the surface, plotted in Fig.~\ref{fig:ratioSurfE} as a function of surface energy,  yield a charge ratio significantly higher at few TeV than those measured by others at surface energies below 300 GeV \cite{compile, l3}.  This rise in the charge ratio at TeV energies is, however, expected as the result of the increasing contribution of kaons to the cosmic ray muon flux at these energies and the greater likelihood for kaons to decay to $\mu^+$  than for pions to decay to $\mu^+$~ \cite{gaisser}.

We use a qualitative model of the charge ratio to show that the rise in the charge ratio at TeV energies seen in Fig.~\ref{fig:ratioSurfE} is consistent with  this expectation.  
Gaisser \cite{gaisser} and Gaisser and Stanev \cite{part} give an expression for the muon intensity at the surface as a function of the muon energy and zenith angle.  It has contributions from both pion and kaon decay and it comes from folding the measured spectrum of cosmic ray primaries with the kinematics of pion and kaon decay.  In our model we assume this expression holds independently for both $\mu^+$ and $\mu^-$.  In addition, this model assumes that charm production can be neglected and that pion and kaon interaction lengths are independent of charge at these energies.  Let $f_{\pi^{+}}$ be the fraction of all pion decays with a detected muon that is positive.  Then $(1 - f_{\pi^+})$ is the fraction of all pion decays with a detected muon that is negative.  Similarly, let  $f_{K^{+}}$ be the fraction of all kaon decays with a detected muon that is positive and $(1 - f_{K^+})$ be the fraction of all kaon decays with a detected muon that is negative.  
The muon charge ratio  $N_{\mu^+}/N_{\mu^-}$ can then be written 
\begin{eqnarray}
\label{phil}
\frac{N_{\mu^+}}{N_{\mu^-}}&& = \left [ 
\frac{f_{\pi^+}}
{1+\frac{1.1 E_{\mu^+}\cos \theta}{\rm 115~GeV}}
  +\frac{0.054f_{K^+}}    {1+\frac{1.1 E_{\mu^+}\cos \theta}{\rm 850~GeV}}
\right ] \Big{/} 
    \nonumber \\ 
& &
\left [ 
\frac{(1 -f_{\pi^+})}{1+\frac{1.1 E_{\mu^-}\cos \theta}{\rm 115~GeV}}
		+  \frac{0.054(1- f_{K^+})}{1+\frac{1.1 E_{\mu^-} \cos \theta}{\rm 850~GeV}}
\right ]~~~~
\end{eqnarray}
The simplest assumption to make in this model is that $f_{\pi^{+}}$ and $f_{K^{+}}$ are independent of energy.   Although this assumption neglects many physical processes that may play a role at these energies, it is an assumption that is a reasonable choice to qualitatively describe our results: a rise in the charge ratio from a plateau at a few hundred MeV to a second higher plateau at a few TeV.  This rise has already been seen in the results from the CORT cosmic ray Monte Carlo~\cite{cort} and the models of Lipari~\cite{lipari2}.  Below we test this simple model with the MINOS data.  

\begin{figure}
\centerline{\epsfig{file=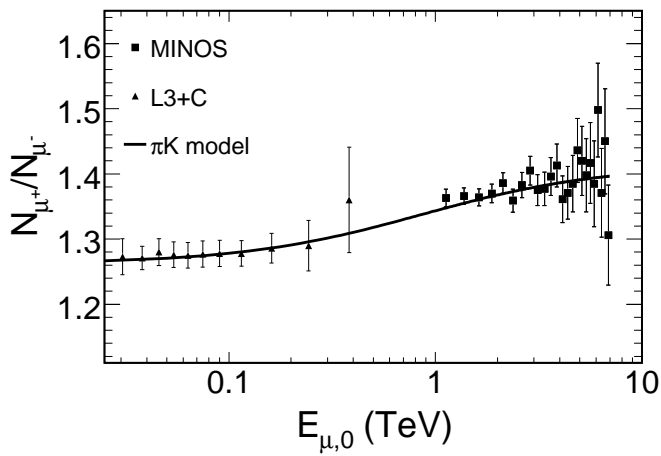,width=3.75in}}
\caption{\label{fig:modelFit}   The $\pi K$ model discussed in the text superposed onto the MINOS and L3+C data sets. The MINOS charge ratio data are from Table~\ref{table:surfCR} with a systematic error of $\pm 0.011$ that has been added in quadrature to the statistical error for each data point.  The L3+C data are taken from \cite{l3}.
}
\end{figure}

We used Eq.~(\ref{phil}) with the MINOS data set and the L3+C data set \cite{l3} to find the values of $f_{\pi^+}$ and $f_{K^+}$ that best describe these two data sets.  We used these data because they have the angular information needed for the fit.  We found the best fit values for $f_{\pi^+}$ and $f_{K^+}$ with a grid search over $(f_{\pi^+}, f_{K^+})$ parameter space.  At each point in the space a $\chi^2$ statistic compared the mean charge ratio weighted by solid angle with the model predictions represented by Eq.~(\ref{phil}). The charge ratio values in each bin had uncertainties given by the quadratic sum of the statistical and systematic errors.  The $\chi^2$ minimimum is found at $f_{\pi^+} = 0.55$ and $f_{K^+} = 0.67$, with $\chi^2/ndf \simeq 1$.  In Fig.~\ref{fig:modelFit}, we have superposed this `$\pi K$' model onto the MINOS and L3+C data sets \cite{l3}.  
The qualitative results of the model are that the observed rise in the muon charge ratio can be explained by the increasing importance of kaon decays to the muon charge ratio as the energy increases from 0.3 -- 1 TeV and that values of $f_{\pi^+}$ and $f_{K^+}$ that are independent of energy are sufficient to describe the MINOS and L3+C data.  

\begin{figure}
\centerline{\epsfig{file= 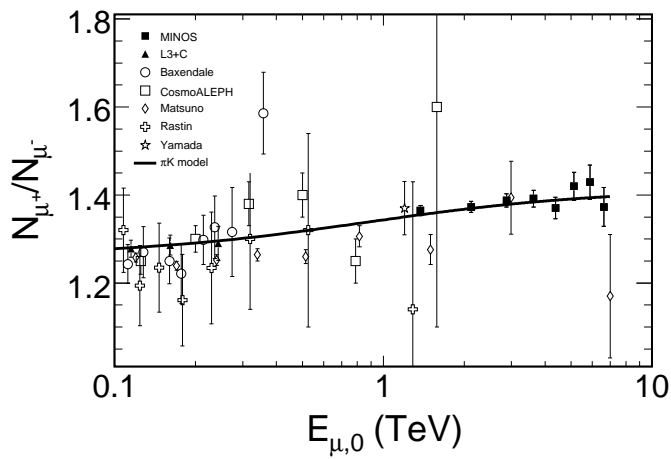,width=3.75in}}
\caption{\label{fig:compilation}   A  compilation of muon charge ratio data from 0.1 to 7 TeV.  The MINOS data have been taken from Table~\ref{table:surfCR}.  Other data: L3+C~\cite{l3}, Baxendale~\cite{bax}, CosmoALEPH~\cite{aleph},  Matsuno~\cite{matsuno}, and Rastin~\cite{rastin}.  The $\pi K$ model is superposed.  
}
\end{figure}

Fig.~\ref{fig:compilation} shows the $\pi K$ model superposed onto a compilation of data from the literature, as well as the MINOS and L3+C data.  The additional data also support the results of this simple model.  

\section{\label{summary}Summary}

The analysis presented here can be separated into three parts.  First, we computed the muon charge ratio underground.  To minimize residual systematic errors, we combined data with both foward and reversed magnetic field running configurations.  When combined, two independent analyses (MIC/BdL) give a muon charge ratio as measured underground of:
\begin{equation}
\label{chargeRatioUnderground}
N_{\mu^+}/N_{\mu^-} = 1.374 \pm 0.004\,({\rm stat.}) ^{+0.012}_{-0.010}\, ({\rm sys.}). \nonumber
\end{equation}
Second, using a map of the Soudan rock overburden, the muon momenta were extrapolated to their corresponding values at the surface, spanning the energy range from 1 to 7 TeV.  Within this range of energies at the surface, the MINOS data are consistent with the charge ratio being energy independent at the two standard deviation level.  The charge ratio as measured by MINOS is significantly higher than measurements by other experiments at surface energies below 300 GeV.  Finally, we used MINOS and L3+C data in a simple model that attributes the rise in the charge ratio with energy
to the increasing contribution of kaon decays.  Fitting the data to the model gives results that are consistent with this picture.  

\begin{center}
{\bf Acknowledgements}
\end{center}

This work was supported by the U.S. Department of Energy, the U.S. National Science Foundation, the U.K. Particle Physics and Astronomy Research Council, the State and University of Minnesota, the Office of Special Accounts for Research Grants of the University of Athens, Greece, FAPESP (Rundacao de Amparo a Pesquisa do Estado de Sao Paulo), CNPq (Conselho Nacional de Desenvolvimento Cientifico e Tecnologico) in Brazil, and the computational resources of the AVIDD cluster at Indiana University.  We gratefully acknowledge the Minnesota Department of Natural Resources for their assistance and for allowing us access to the facilities of the Soudan Underground Mine State Park.  We also thank the crew of the Soudan Underground Physics laboratory for their tireless work in building and operating the MINOS detector.  We acknowledge G. Bodwin of Argonne for useful theoretical discussions.

\bibliography{muPaperLast}

\end{document}